\documentclass[preprint2]{aastex6}

\usepackage{graphicx,fancyhdr,rotating,amsmath,epsf,txfonts,epstopdf,multirow}
\def \kms {{\rm km\;s$^{-1}$}}

\def \arcsec {$^{''}$}
\def \siiv {Si\,{\sc iv}}
\def \cii {C\,{\sc ii}}
\def \mgiik {Mg\,{\sc ii}\,k}
\def \mgiih {Mg\,{\sc ii}\,h}

\begin{document}
\title{Narrow-line-width UV bursts in the transition region above Sunspots observed by IRIS}
\author{
\sc{Zhenyong Hou\altaffilmark{1}, Zhenghua Huang\altaffilmark{1}, Lidong Xia\altaffilmark{1}, Bo Li\altaffilmark{1}, Maria S. Madjarska\altaffilmark{2,1,3}, Hui Fu\altaffilmark{1}, Chaozhou Mou\altaffilmark{1}, Haixia Xie\altaffilmark{1}}
}
\altaffiltext{1}{Shandong Provincial Key Laboratory of Optical Astronomy and Solar-Terrestrial Environment, Institute of Space Sciences,
Shandong University, Weihai, 264209 Shandong, China; {\it z.huang@sdu.edu.cn, xld@sdu.edu.cn}}
\altaffiltext{2}{Armagh Observatory, College Hill, Armagh BT61 9DG, N. Ireland, UK}
\altaffiltext{3}{Max Planck Institute for Solar System Research, Justus-von-Liebig-Weg 3, 37077, G\"ottingen, Germany}

\begin{abstract}
Various small-scale structures abound in the solar atmosphere above active regions,
    playing an important role in the dynamics and evolution therein.
We report on a new class of small-scale transition region structures in active regions,
    characterized by strong emissions but extremely narrow \siiv\ line profiles
    as found in observations taken with the Interface Region Imaging Spectrograph (IRIS).
Tentatively named as Narrow-line-width UV bursts (NUBs),
    these structures are located above sunspots and comprise of one or multiple compact bright cores
    at sub-arcsecond scales.
We found six NUBs in two datasets (a raster and a sit-and-stare dataset).
Among these, four events are short-living with a duration of $\sim$10\,mins while two last for more than 36\,mins.
All NUBs have Doppler shifts of 15--18\,\kms, while the NUB found in sit-and-stare data possesses an additional component
   at $\sim$50\,\kms\ found only in the \cii\ and Mg\,{\sc ii} lines.
Given that these events are found to play a role in the local dynamics,
    it is important to further investigate the physical mechanisms
    that generate these phenomena and their role in the mass transport in sunspots.
\end{abstract}

\maketitle

\section{Introduction}
\label{sect_intro}

Small-scale structures and activities abound in solar active regions (ARs),
    including explosive events\,\citep{1989SoPh..123...41D, 1997Natur.386..811I, 2014ApJ...797...88H},
    blinkers\,\citep{1999A&A...351.1115H, 2002SoPh..206..249P},
    small-scale loops\,\citep{2015ApJ...810...46H},
    spicules and fibrils\,\citep{2000A&A...360..351W,2012SSRv..169..181T,2014ApJ...792L..15P,2015ApJ...799L...3R}, to name a few.
Thanks to recent high-resolution observations of ARs,
    activities have been seen on even finer scales.
For example, observations made with the high-Resolution Coronal Imager\,\citep[Hi-C,][]{2014SoPh..289.4393K}
    have indicated signatures of magnetic braids with scales of $\sim0.2$\arcsec\ in AR loops\,\citep{2013Natur.493..501C},
    heating events with a similar size in both moss\,\citep{2013ApJ...770L...1T}
        and inter-moss loops\,\citep{2013ApJ...771...21W},
    as well as fine loops at sub-arcsecond scales\,\citep{2013A&A...556A.104P}.
Using Hi-C observations, \citet{2014ApJ...784..134R} reported on the discovery of
     small-scale brightenings named as ``EUV bright dots'' (EUV BDs) found at the base of large-scale loops rooted at the edge of ARs
     with a characteristic duration of 25\,s and a typical length of $\lesssim1$\arcsec.
More recently, the Interface Region Imaging Spectrograph\,\citep[IRIS,][]{2014SoPh..289.2733D}
     has discovered very broad \siiv\ emission profiles blended by strong absorption lines
     in small-scale ($\sim1$\arcsec) compact brightenings in ARs~\citep{2014Sci...346C.315P}.
These brightenings, named ``hot bombs'', were suggested to be signatures of hot plasmas in the photosphere generated by magnetic reconnections.
In addition, \citet{2014ApJ...790L..29T} reported on sub-arcsecond BDs implicating heating events
     in the transition region (TR) above umbrae and penumbrae of sunspots underlying ARs.
\citet{2016arXiv160700306D} found that these BDs do not have a chromospheric response and suggested their TR formation.
   Similarly, BDs in sunspots were also recently reported in Hi-C observations\,\citep{2016ApJ...822...35A}.
In sunspot umbrae and penumbrae, \citet{2014ApJ...789L..42K} presented IRIS observations
     of small-scale brightenings associated with supersonic downflows,
     which were suggested to play a role in heating the TR above sunspots.
Penumbral jets have been investigated with IRIS data and were found to be heated to TR temperature\,\citep{2015ApJ...811L..33V}.

\par
In the present work, we report on a new class of fine-scale structures in ARs,
     characterized by very bright emissions but extremely narrow profiles in the \siiv\,line in the TR above sunspots as observed by IRIS.
In what follows, we present our observations in Section\,\ref{sect_obs},
     describe the results in Section\,\ref{sect_result},
     and summarize our findings in Section\,\ref{sect_conclusion} where
     some discussions are also offered.

\begin{table*}[!ht]
  \centering
  \caption{\textbf{Parameters of the bright features.}}\label{table}
  \renewcommand{\multirowsetup}{\centering}
  \renewcommand{\arraystretch}{0.7}
  \begin{tabular}{c c c c c c c c c c c}
  \\\hline
ID&Lifetime\tablenotemark{a}&channel&Intensity&R$_{1}$\tablenotemark{b}&FWHM\tablenotemark{c}&R$_{2}$\tablenotemark{d}&V$_{nt.}$&V$_{Dopp}$&Area&I$_{1394}$/I$_{1403}$ \\
&[mins]&[\AA]&[DN/s]&&[m\AA]&&[\kms]&[\kms]&[arcsec$^{2}$]&\\
\hline
  \multirow{6}{10mm}{1} & \multirow{6}{10mm}{$\sim$\,6}   & \siiv\ 1394  & 727   & 5.0    & 65.9   & 0.3 &5.6$\pm$0.8 & 16.0$\pm$1.1 & 6.4 & \multirow{6}{10mm}{1.9} \\
                             &                            & \siiv\ 1403  & 380   & 5.0    & 66.3   & 0.4 &5.6$\pm$0.7 & 16.2$\pm$1.0 & ---  \\ \cline{3-10}
                             &                            & \cii\ 1334   & 474   & 7.9    & 98.4   & 0.5 &---         & ---          & ---  \\
                             &                            & \cii\ 1336   & 672   & 8.3    & 104.1  & 0.5 &---         & ---          & ---  \\ \cline{3-10}
                             &                            & \mgiik\ 2796 & 1634  & 1.1    & 273.5  & 0.6 &---         & ---          & ---  \\
                             &                            & \mgiih\ 2803 & 1319  & 1.2    & 246.7  & 0.6 &---         & ---          & ---  \\ \hline
  \multirow{6}{10mm}{2} & \multirow{6}{10mm}{$>$36}       & \siiv\ 1394  & 579   & 4.0    & 72.5   & 0.4 &6.8$\pm$0.8 & 15.1$\pm$1.4 & 36.3& \multirow{6}{10mm}{1.7}\\
                             &                            & \siiv\ 1403  & 347   & 4.6    & 66.0   & 0.3 &5.6$\pm$0.5 & 15.9$\pm$1.5 & ---  \\ \cline{3-10}
                             &                            & \cii\ 1334   & 12    & 0.2    & 107.7  & 0.6 &---         & ---          & ---  \\
                             &                            & \cii\ 1336   & 16    & 0.2    & 91.3   & 0.4 &---         & ---          & ---  \\ \cline{3-10}
                             &                            & \mgiik\ 2796 & 392   & 0.3    & 231.9  & 0.5 &---         & ---          & ---  \\
                             &                            & \mgiih\ 2803 & 282   & 0.3    & 213.4  & 0.5 &---         & ---          & ---  \\ \hline
  \multirow{6}{10mm}{3} & \multirow{6}{10mm}{$\sim$\,3.5} & \siiv\ 1394  & 2138  & 15.2   & 85.4   & 0.5 &7.9$\pm$1.3 & 17.0$\pm$2.7 & 16.9& \multirow{6}{10mm}{1.9} \\
                             &                            & \siiv\ 1403  & 1151  & 15.9   & 80.7   & 0.4 &7.5$\pm$1.3 & 17.2$\pm$2.7 & ---  \\ \cline{3-10}
                             &                            & \cii\ 1334   & 300   & 5.0    & 134.5  & 0.7 &---         & ---          & ---  \\
                             &                            & \cii\ 1336   & 422   & 5.2    & 155.5  & 0.7 &---         & ---          & ---  \\ \cline{3-10}
                             &                            & \mgiik\ 2796 & 3144  & 2.1    & 455.2  & 1.0 &---         & ---          & ---  \\
                             &                            & \mgiih\ 2803 & 2625  & 2.4    & 421.5  & 1.0 &---         & ---          & ---  \\ \hline
  \multirow{6}{10mm}{4} & \multirow{6}{10mm}{$\sim$\,10}  & \siiv\ 1394  & 2754  & 19.6   & 88.8   & 0.5 &8.7$\pm$0.7 & 16.9$\pm$3.5 & 7.5 & \multirow{6}{10mm}{1.9} \\
                             &                            & \siiv\ 1403  & 1470  & 20.3   & 86.6   & 0.5 &8.4$\pm$0.9 & 16.9$\pm$3.4 & ---  \\ \cline{3-10}
                             &                            & \cii\ 1334   & 608   & 10.1   & 118.0  & 0.6 &---         & ---          & ---  \\
                             &                            & \cii\ 1336   & 844   & 10.4   & 138.5  & 0.6 &---         & ---          & ---  \\ \cline{3-10}
                             &                            & \mgiik\ 2796 & 2510  & 1.7    & 410.8  & 0.9 &---         & ---          & ---  \\
                             &                            & \mgiih\ 2803 & 2058  & 1.9    & 371.4  & 0.9 &---         & ---          & ---  \\ \hline
  \multirow{6}{10mm}{5} & \multirow{6}{10mm}{$>$45}       & \siiv\ 1394  & 682   & 4.8    & 84.5   & 0.5 &8.4$\pm$0.8 & 15.4$\pm$1.8 & 23.0& \multirow{6}{10mm}{1.8} \\
                             &                            & \siiv\ 1403  & 370   & 5.1    & 84.2   & 0.5 &8.3$\pm$0.9 & 15.6$\pm$1.9 & ---  \\ \cline{3-10}
                             &                            & \cii\ 1334   & 159   & 2.7    & 128.0  & 0.7 &---         & ---          & ---  \\
                             &                            & \cii\ 1336   & 210   & 2.6    & 155.5  & 0.7 &---         & ---          & ---  \\ \cline{3-10}
                             &                            & \mgiik\ 2796 & 3104  & 2.0    & 414.1  & 0.9 &---         & ---          & ---  \\
                             &                            & \mgiih\ 2803 & 2501  & 2.3    & 381.7  & 0.9 &---         & ---          & ---  \\ \hline
  \multirow{6}{10mm}{6} & \multirow{6}{10mm}{$\sim$\,14}  & \siiv\ 1394  & 158   & 9.1    & 56.8   & 0.4 &3.6$\pm$1.0 & 16.7$\pm$2.2 & 1.0& \multirow{6}{10mm}{2.0}  \\
                             &                            & \siiv\ 1403  & 79    & 8.8    & 61.3   & 0.4 &4.4$\pm$1.2 & 18.0$\pm$2.2 & ---  \\ \cline{3-10}
                             &                            & \cii\ 1334   & 58    & 3.8    & 159.3  & 1.1 &---         & ---          & ---  \\
                             &                            & \cii\ 1336   & 91    & 4.1    & 207.5  & 1.2 &---         & ---          & ---  \\ \cline{3-10}
                             &                            & \mgiik\ 2796 & 469   & 0.9    & 455.0  & 1.1 &---         & ---          & ---  \\
                             &                            & \mgiih\ 2803 & 285   & 0.8    & 341.7  & 1.0 &---         & ---          & ---  \\ \hline
\end{tabular}
\tablenotetext{a}{The lifetime determined from the IRIS 1400\,\AA\ SJIs. Events\,2 and 5 are both visible during the entire observing period.}
\tablenotetext{b}{The intensity ratios of the events to the reference spectra.}
\tablenotetext{c}{Full width at half maximum of the line profiles subtracted by the instrumental width (31.8\,m\AA\ for \siiv, 28.6\,m\AA\ for \cii\ and 50.54\,m\AA\ for Mg\,{\sc ii}).}
\tablenotetext{d}{The ratios of the FWHMs of the events to the reference spectra.}
\end{table*}

\section{OBSERVATIONS}
\label{sect_obs}
\begin{figure*}[!ht]
\centering
\includegraphics[trim=0cm 0.7cm 0cm 3.5cm,width=0.85\textwidth]{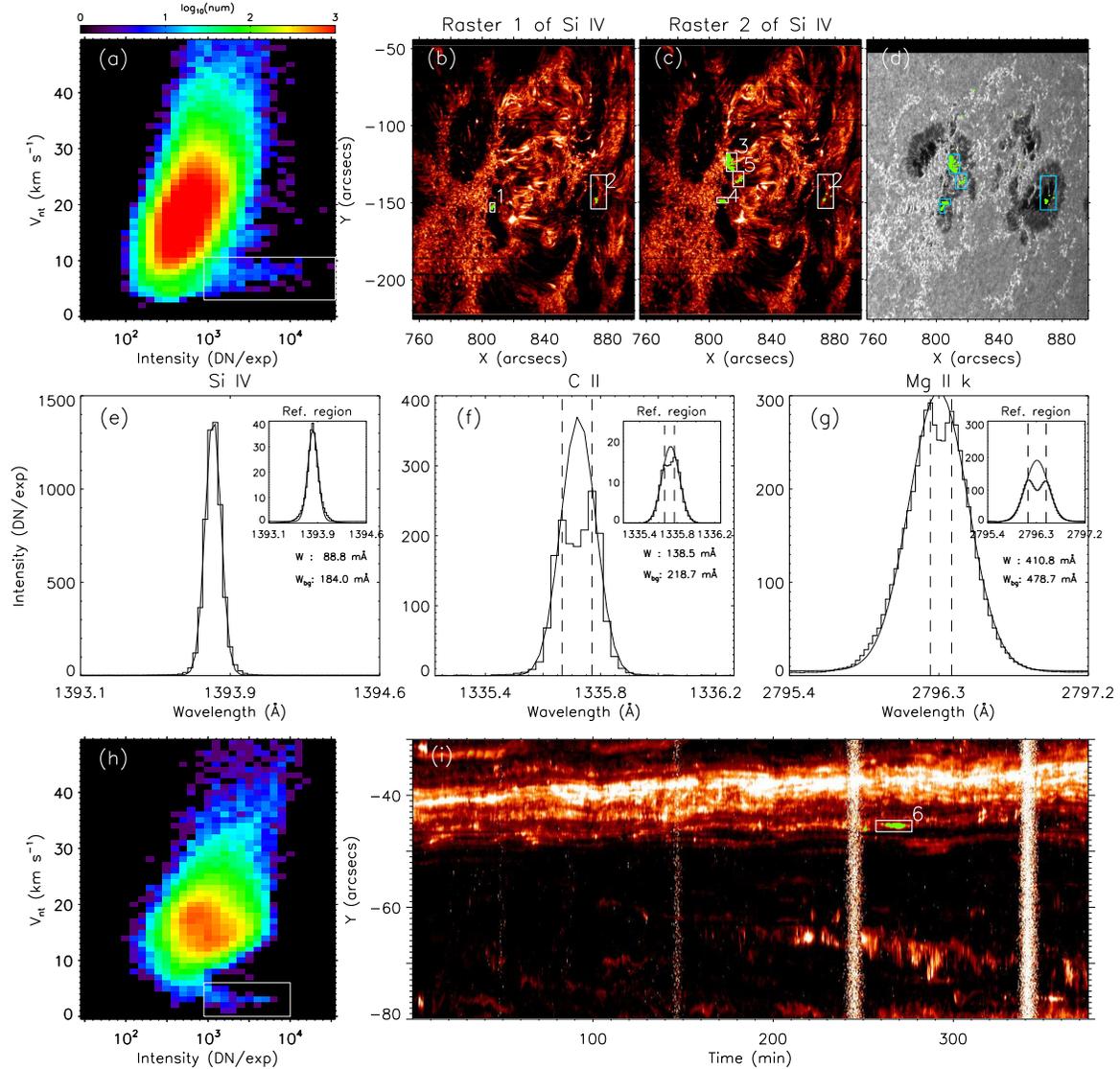}
\caption{
  Identification of the events in the observed FOVs.
  Panel\,(a): two-dimensional histogram of the non-thermal velocities and intensities obtained from DATA1.
  The solid-line square marks the tail of the histogram that was used to make the identification of the events.
  Panels\,(b) and (c) present the \siiv\ radiance images of the two rasters of the same region with the identifications of the events superimposed (green symbols).
  Panel\,(d) displays the same region with the identified events observed in the continuum near 2832\,\AA\ to give an overview of the photosphere of the region.
  Panels\,(e)--(g) give example spectra with single Gaussian fits (solid lines) averaged from event\,4.
  The dashed vertical lines (in panels\,f and g) mark the wavelength range that shows absorption dips and are excluded from the Gaussian fits.
  Panels\,(f) and (i) present the results from DATA2.
}
\label{f1}
\end{figure*}

The IRIS observations analyzed in this study include two datasets.
The first (DATA1) is a raster dataset taken on 2014 February 16 from 20:19\,UT to 21:04\,UT when IRIS was targeting AR 11974
    and scanned  the same area twice with a 0.35\arcsec\-wide slit and an exposure time of 2\,s.
The spectral slit scanned a 140\arcsec\,$\times$\,175\arcsec\ field-of-view (FOV) with a step size of 0.35\arcsec\ and an along-slit pixel size of 0.17\arcsec,
in which several sunspots and their surrounding plages were observed (see Figs.\,\ref{f1}b--d).

The second dataset (DATA2) was taken on 2014 March 10 from 04:10\,UT to 10:26\,UT, when
  IRIS targeted NOAA\,11998 in a sit-and-stare mode with an exposure time of 15\,s.
In both datasets, we analyzed the spectral data taken in the \siiv\,1394\,\AA\ and 1403\,\AA,
  \cii\,1334\,\AA\ and 1336\,\AA,
  \mgiik\,2796\,\AA\ and Mg\,{\sc ii}\,h\,2803\,\AA\ lines
   as well as the continuum around 2832\,\AA.
In both datasets, the slit-jaw (SJ) data with a spatial resolution of 0.34\arcsec taken in the 1330\,\AA\ and 1400\,\AA\ channels
   were analyzed. The cadence of these images for DATA1 and DATA2 was 14\,s and 33\,s, respectively.

\par
In order to determine the rest wavelength of the \siiv\ line,
   we used the chromospheric Fe\,{\sc ii} 1392.82\,\AA\ line as a reference to zero Doppler-shift,
   and the vacuum wavelengths of the \siiv\,1394\,\AA\ and 1403\,\AA\ are taken as 1393.76\,\AA\ and
   1402.77\,\AA\ (see the IRIS Technical Note 20 for details).
The residual orbital variation has been removed by the new version of the procedure \textit{iris\_orbitvar\_corr\_l2.pro}.

\begin{figure*}[!ht]
\centering
\includegraphics[trim=1.5cm 0.6cm 0cm 1.5cm,width=0.9\textwidth]{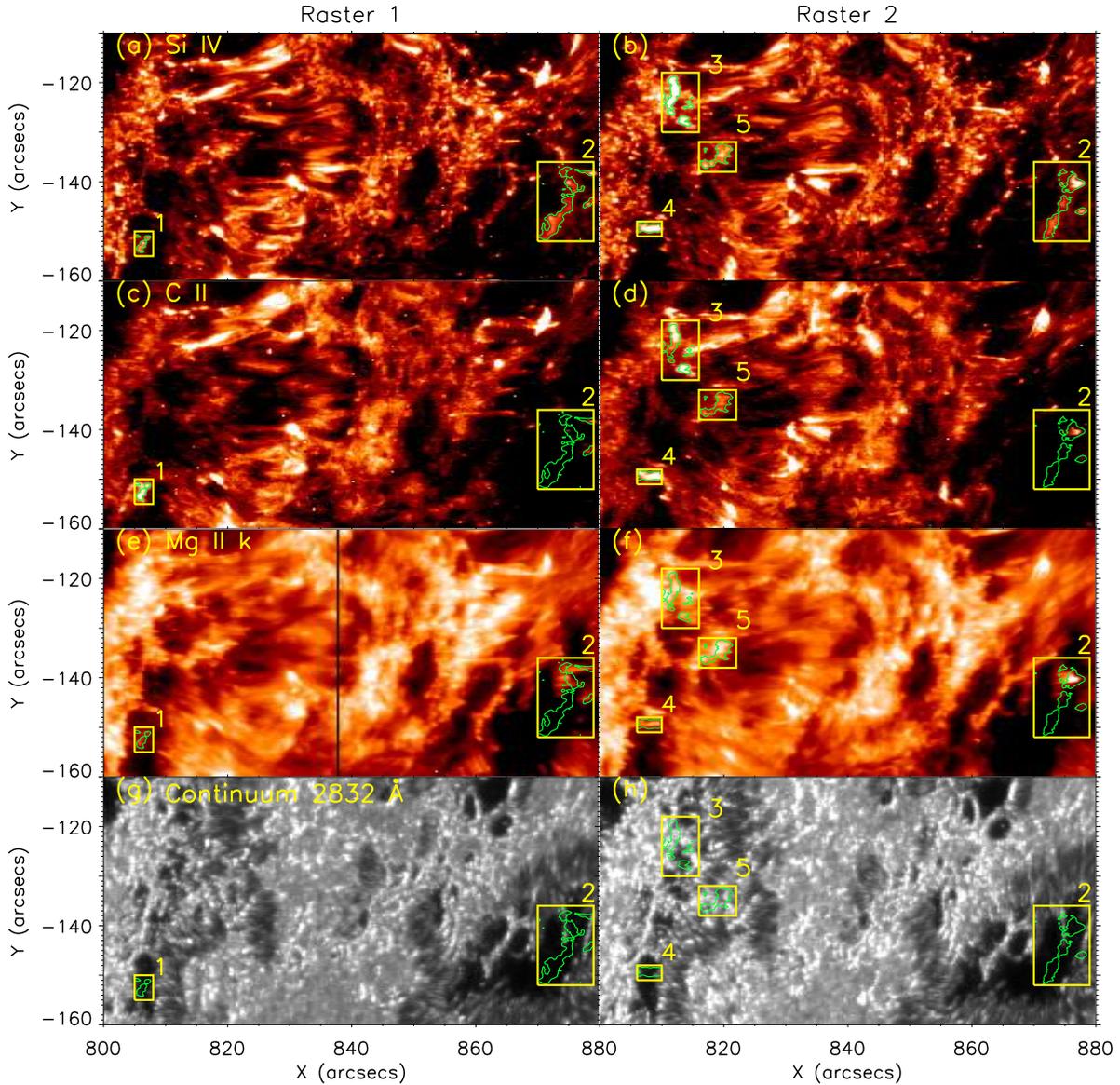}
\caption{
  Zoom-in view of the events in the \siiv\ (a\&b), \cii\ (c\&d), \mgiik\ (e\&f) and continuum 2832\,\AA\ (g\&h) radiance images in DATA1.
  The contour (green lines) taken from the \siiv\ radiance images (only the areas around the events are considered) are superimposed on all images taken from the same raster to illustrate the locations of the events.
}
\label{f2}
\end{figure*}

\begin{figure*}[!ht]
\centering
\includegraphics[trim=0cm 0.3cm 0cm 0.4cm,width=\textwidth]{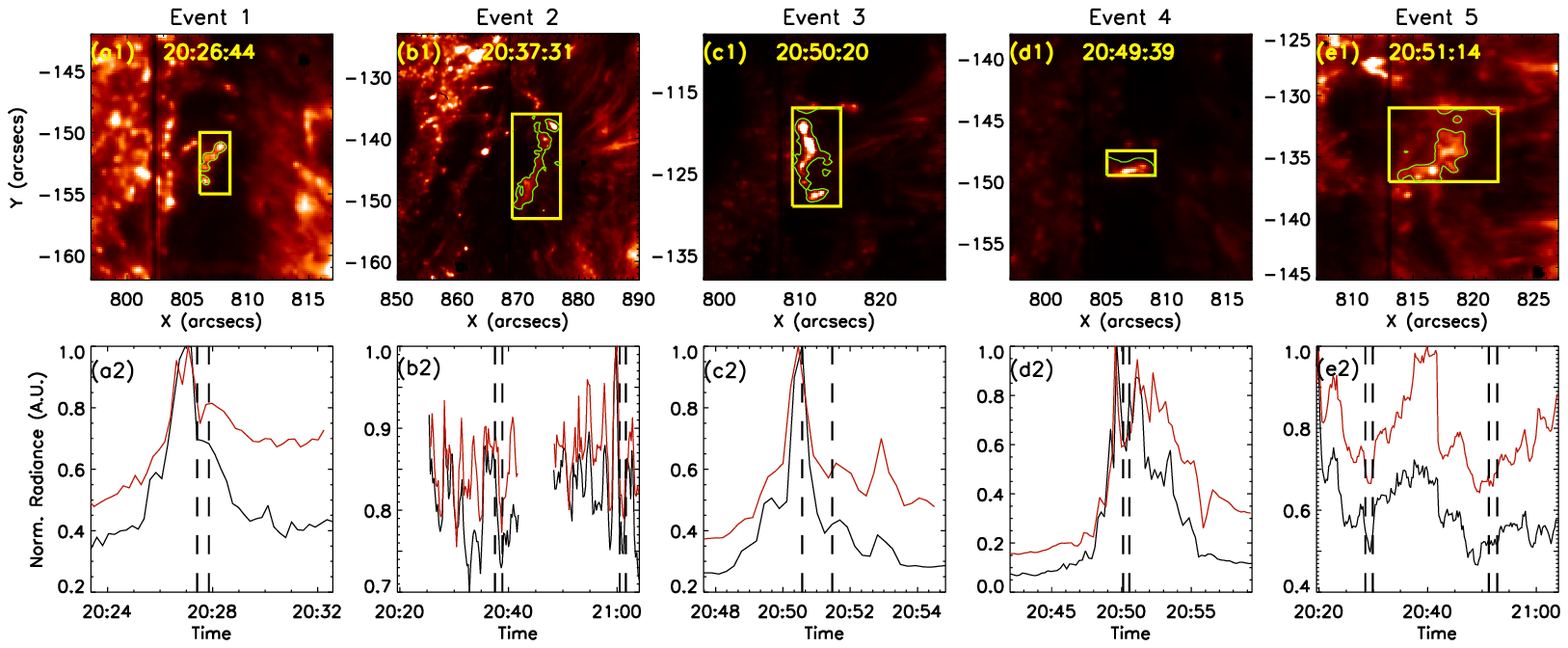}
\caption{
  Events\,1--5 identified in DATA1 seen in IRIS SJ 1400\,\AA\ (Panels\,a1--e1).
  They are marked by the solid-line squares (yellow) and contoured by green lines.
  Their IRIS SJ 1400 and 1330\,\AA\ lightcurves are given as black and red lines in Panels\,a2--e2.
  The dashed lines denote the period of the IRIS spectral slit scan. An animation is given online.}
\label{f3}
\end{figure*}

\section{Results}
\label{sect_result}
\subsection{The raster dataset}
We started our analysis by deriving the \siiv\,1394\,\AA\ line widths in the observed FOV of DATA1.
To this end, we applied a single-Gaussian fit to all the spectra with good signal-to-noise ratios.
We excluded those pixels with intensity at line center is lower than that 30 times of the background (taken in the continuum at $\sim$200\,\kms\ Doppler shift),
    and those pixels that are spiked by high-energy particles and cosmic rays.
The non-thermal velocities were computed by using the instrumental broadening
    obtained by the pre-flight measurements (31.8\,m\AA\ for \siiv) and
    by assuming a formation temperature of $T_\mathrm{max}=6.3\times10^4$~K
    as given by CHIANTI\,\citep[V7.1.3,][]{1997A&AS..125..149D,2013ApJ...763...86L}.
We then produced a two-dimensional histogram of intensities and non-thermal velocities
   for the two spectral scans (see Fig.\,\ref{f1}a).
The \siiv\ non-thermal velocities in the ARs
    range from 3\,\kms\ to 50\,\kms\ with the majority of the pixels lying in the range between 10 and 30\,\kms,
   in agreement with previous studies\,\citep[e.g.][etc.]{1993SoPh..144..217D,1999A&A...349..636T,2015ApJ...799L..12D}.
In addition, Fig.~\ref{f1}a indicates that
   despite the spread, there exists a positive correlation between the non-thermal velocities and intensities.
A similar tendency has been reported by \citet{1998ApJ...505..957C} who studied a quiet-Sun region using SUMER observations.

\par
From the histogram, a tail in the distribution is clearly visible as denoted by the white square in Fig.\,\ref{f1}a,
    corresponding to relatively strong emissions but very small non-thermal velocities (i.e. line widths).
It turns out that the locations of the points in this tail,
    as given by the green pixels in the raster images (Figs.\,\ref{f1}b to \ref{f1}d),
    are clustered around a few specific positions.
To better see them, we enclosed these green pixels with solid-line squares
    and labeled them events\,1 to 5 in Figs.\,\ref{f1}b--d.
An example of the \siiv, \cii\ and \mgiik\ line profiles
    averaged over the whole area of event\,4 is given in Figs.\,\ref{f1}e--g.
For reference, the profile averaged over the entire FOV is also presented (the insets).
From Fig.~\ref{f1}e one sees that the average line width of the \siiv\,1394\,\AA\ profile for Event\,4, for instance, is 88.8\,m\AA\
    (or equivalently 8.7$\pm$0.7\,\kms\ in terms of non-thermal velocities),
    substantially narrower than the reference profile for which a width of
    184.0\,m\AA\ (or 23.8\,\kms) is derived.
In contrast, its intensity is stronger by a factor of $\sim 20$ (see also Table\,\ref{table}).
For comparison, we note that the line width and integrated intensity of the \siiv\,1394\,\AA\ line in the sunspots
    are 64.6\,m\AA\ (or 5.6\,\kms) and 45.0\,DN/s, respectively. The line width (non-thermal velocity) measured in the sunspot is close to that of the event, while the intensity is much smaller (the intensity of Event 4 is about 2750\,DN/s, see Table\,\ref{table}).
In Figs.~\ref{f1}f and \ref{f1}g, the \cii\ and \mgiik\ line profiles are fitted by excluding the self-reversal portion\,\citep{xiathesis}.
One sees that the \cii\ profile for event 4 behaves in a similar fashion to \siiv\,
    even though the \mgiik\ profile is only marginally narrower than the reference spectrum.
The Doppler velocities of events 1--5 range from 15 to 17\,\kms\
measured in the two  \siiv\ lines.

\par
Figure~\ref{f2} presents a zoom-in view of these bright features on the images
    taken in the \siiv, \cii, \mgiik\ and continuum 2832\,\AA\ lines.
All pixels in events 1, 3 and 4 are characterized by a narrow profile together with some strong emission in \siiv.
However, for the two long-duration events (events 2 and 5),
    only a fraction of the pixels correspond to strong emissions and narrow line widths,
    thereby falling into the tail of the histogram in Fig.~\ref{f1}a.
While the profiles at the rest pixels are significantly narrower than the reference spectra,
    the intensities are not as strong.
Consider Figs.~\ref{f2}c to \ref{f2}f.
One sees that events\,1, 3, 4 and 5 can be clearly seen in both the \cii\ and \mgiik\ radiance images,
    whereas event 2 can be hardly discerned.
Regarding the continuum 2832\,\AA\ radiance images (Figs.~\ref{f2}g and \ref{f2}h),
    one can see that all these five events are associated with sunspots.
In particular, events 1 and 4 are located above the umbra of one of the small sunspots in the region,
    while events\,3 and 5 appear above some brighter regions (likely the penumbra of the sunspot).
On the other hand, event\,2 is located above the umbra of a neighboring sunspot,
    which formed around February 13.
We found that the \siiv\ line widths of these events are close to that of the spectra averaged from the dark sunspot region (64.6\,m\AA),
    suggesting that the plasmas in these events are probably linked to the sunspot.

\par
These events can also be examined from the perspective of their temporal evolution.
For this purpose we attached an online animation to Fig.\,\ref{f3}, showing
    these events in the IRIS 1400\,\AA\ SJ images (SJIs).
In addition, snapshots for individual events are also presented in the top row of Fig.~\ref{f3}.
The lightcurves for these events in both SJ 1330\,\AA\ and 1400\,\AA\
    are shown in the bottom row of Fig.~\ref{f3} (see the red and black curves),
    from which one sees that they are similar in both channels as far as
    their qualitative behavior is concerned.
Interestingly, events 1, 3, and 4 tend to show up in their lightcurves as
    a short-lived spike with a duration of $\lesssim 10$\,min.
The spectral slit scanned events\,1 and 3 in their decaying phase,
    but traversed event 4 at its peak.
Events 2 and 5 show a different behavior in that they actually brightened multiple times,
    as evidenced by the many spikes in their lightcurves.
These two events have a longer lifetime of $\gtrsim 36$\,min.
A closer inspection further indicates that these events
    are elongated structures and comprise more than one bright cores with scales of $\lesssim 1$\arcsec.
For each individual event, the bright cores tend to flare up simultaneously, and they last from 41 to 297\,s with an average of 146\,s.

\par
Table\,\ref{table} lists the parameters for all events,
    giving their lifetimes,
    integrated intensities,
    intensity ratios to the reference region,
    full widths at half maximum (FWHMs),
    ratios of their FWHMs to the reference spectra,
    their non-thermal velocities,
    Doppler velocities,
    and areas,
    as well as the associated intensity ratios of \siiv\ 1394\,\AA\ to \siiv\ 1403\,\AA.
Note that event 6 is what we identified in the sit-and-stare observation (see section~\ref{sect_sns}).
    All these strongly emitting events correspond to
    non-thermal velocities $\lesssim 10$\,\kms\ (see Table\,\ref{table}).
They are all small-scale features and their areas in the \siiv\ raster images are
    in the range of $6\sim36$\,arcsec$^{2}$.
Furthermore, the associated intensity ratios
    of \siiv\,1394\,\AA\ to 1403\,\AA\ range from $1.7$ to $2.0$.
For reference, the intensity ratio of the \siiv\ doublet measured
    in the reference spectrum is $\sim 1.93$.
All these values are close to the ratio of a transition probability of 2\,\citep{1993SoPh..144..217D,1998ApJ...505..957C},
    indicative of an optically thin line formation.

The O\,{\sc iv}\,1401.16\,\AA\ is detected in a part of events 2, 3 and 4,
   while the O\,{\sc iv} 1399.77\,\AA\ appears only in a part of event 2.
The ratio of these two lines in event\,2 indicates an electron density of 1.5$\times10^{10}$\,cm$^{-3}$.

In these events, we found various types of Mg\,{\sc ii}\,triplet profiles (near 2791.6\,\AA\ and 2798.8\,\AA) without any typical behavior.
None of them is comparable with those shown in \citet{2015ApJ...806...14P}.
We, therefore, do not pursue any further analysis.

We found that a weak response in the AIA~304~\AA, 171~\AA, and 211~\AA\ channels is associated with these events.
We believe that this response is caused by the TR emission contamination of the coronal channels\,\citep[see][for more details]{2010A&A...521A..21O}.
We will investigate this issue in a future dedicated study using simultaneous EIS and IRIS data.

\begin{figure*}[!ht]
\centering
\includegraphics[trim=0cm 0.3cm 0cm 0.3cm,width=\textwidth]{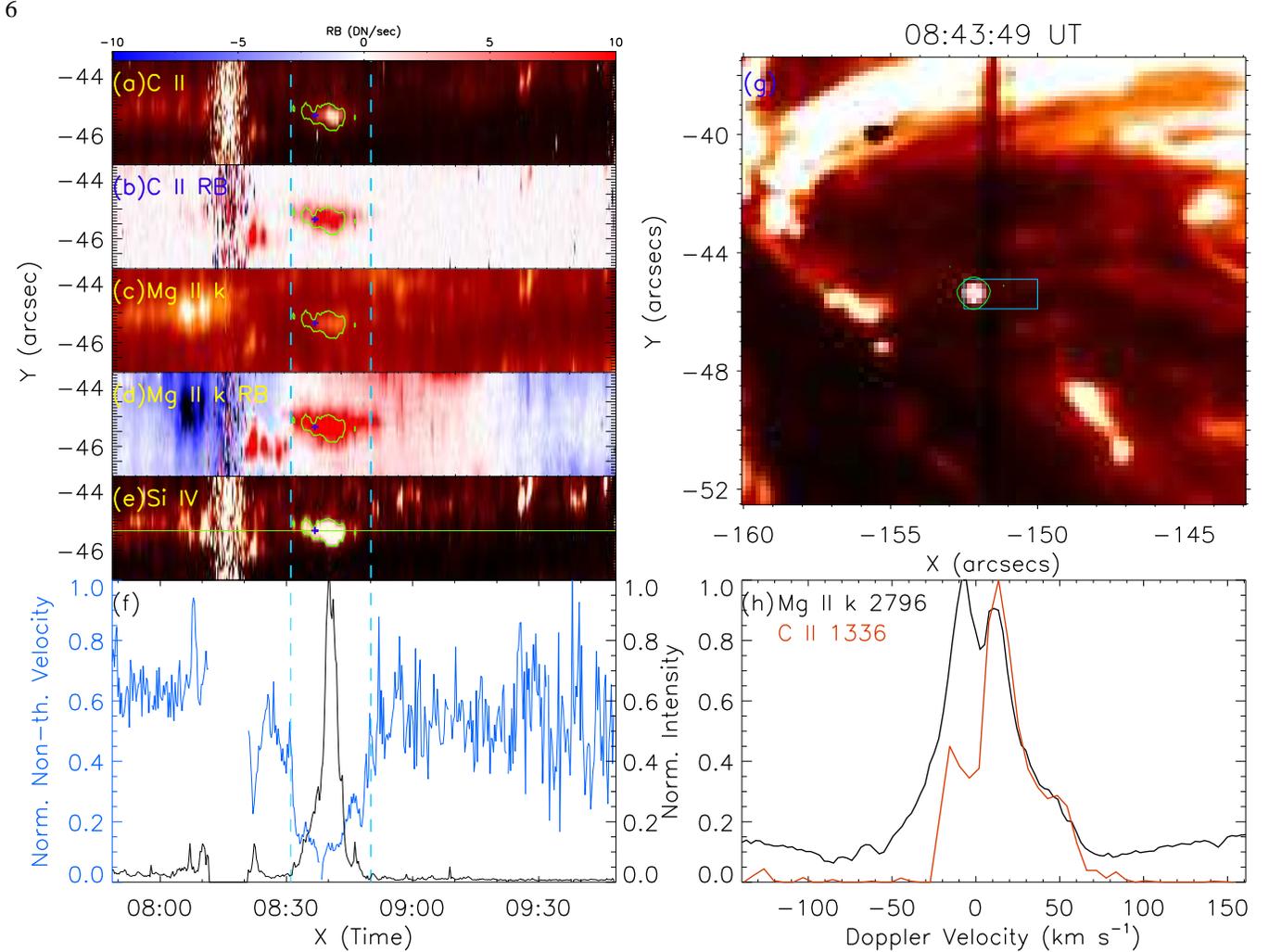}
\caption{
  Evolution of event 6 (in DATA2) seen in \cii\ (a), \cii\ RB asymmetry at 45--55\,\kms\ (b),
    \mgiik\ (c), \mgiik\ RB asymmetry at 45-55\,\kms\ (d),
    \siiv\ (e) and SJ 1400\,\AA\ (g).
  The temporal variation of the \siiv\ intensity and line width are shown with black and blue lines in (f), respectively.
  The dashed lines denote the period when the event occurred.
  Example profiles (taken from the location marked by a plus symbol in panel a--e) of \cii\ (red) and \mgiik\ (black) are shown in (h).
  An animation is given online.}
\label{f4}
\end{figure*}
\subsection{The sit-and-stare sequence}
\label{sect_sns}
We also analyzed a sit-and-stare sequence, in which an event with very small line widths and strong intensities is observed.
The two-dimensional histogram and the \siiv\ radiance image are given in Figs.\,\ref{f1}h and \,\ref{f1}i.
From the histogram, one readily discerns a tail marked by the white square in Fig.\,\ref{f1}h,
    which corresponds to a brightening event labeled `6' in Fig.\,\ref{f1}i.

In Fig.\,\ref{f4} and the online animation, we present the temporal variation of event\,6 seen in \cii, \siiv, \mgiik\
    and SJ 1400\,\AA\ (Fig.\,\ref{f4}).
The event appears as a compact brightening and lasts for $\sim 14$\,mins,
    having a size of 1.0\,arcsec$^{2}$ (determined from the SJ 1400\,\AA\ image).
The temporal variation of \siiv\ line width and intensity are given in Fig.\,\ref{f4}f,
    from which one sees a dramatic decrease in line-width accompanying a drastic increase in intensity.
The line width of this event decreases by $\sim80\%$ while the intensity increases by a factor of $\sim20$ (see Fig.\,\ref{f4}f). Note that the R$_1$ and R$_2$ of this event listed in Table\,\ref{table} were given based on the reference spectra averaging from the entire FOV.
This event is found to be located above the penumbra of a sunspot seen in the IRIS SJ 2832\,\AA.
Again, the intensity ratio of the \siiv\ doublet is $2$ (see Table\,\ref{table}),
    suggesting optically thin emission.

In event\,6, the \siiv\ spectra are regularly Gaussian,
    and the Doppler shifts are measured as $\sim 17$\,\kms\ (red-shifted).
An interesting characteristic of event\,6 is that
    the \cii\ and \mgiik\ show a clear enhancement in the red-wing at about 50\,\kms\ (see an example given in Fig\,\ref{f4}h and more in the online animation).
The red-wing enhancement can be better demonstrated by the \cii\ and \mgiik\ red-blue asymmetry
    \citep{2009ApJ...701L...1D,2011ApJ...738...18T} of the region at 45--55\,\kms\ (see Figs.\,\ref{f4}b and d).
This indicates a supersonic downflow in the chromosphere.
We note that supersonic downflows above sunspots have also been seen in IRIS observations\,\citep{2014ApJ...789L..42K, 2016ApJ...821L..30K}.
However, in contrast to the narrow line profiles,
    these supersonic flows reported previously show broad TR spectra.

\section{Discussion and conclusions}
\label{sect_conclusion}

In this work we identified six events with strong emissions but narrow spectral profiles
   in the IRIS \siiv\ spectral data obtained in ARs.
These events were found to occur in compact regions in the corresponding radiance images,
    and are therefore named as Narrow-line-width UV bursts (NUBs).
We identified five events in raster data and one in a sit-and-stare sequence.
These phenomena appear as one or a group of sub-arcsecond compact brightenings, either in the form of some short-lived bursts with a duration of $\sim 10$\,mins (events\,1, 3, 4 and 6) or as a series of brightenings lasting for $\sim 30$\,mins (events\,2 and 5). We suggest that the phenomena studied here  are of the same class given their similar characteristics. The different appearance from event to event is possibly due to different plasma conditions and magnetic topology. To make an appropriate classification, further investigations are required.
The intensity ratios of the \siiv\ doublet are close to $2$, indicating optically-thin line formation.
The Doppler velocities range from 15 to 18\,\kms.
Compared to the reference spectra, the NUBs possess only about half of the line width and emissions enhanced by a factor of $4-20$.
The average non-thermal velocities of these events range from 3.6 to 8.7\,\kms\ in \siiv\,
   which are significantly smaller than the reference non-thermal velocity of $\sim$22\,\kms\ from DATA1 and DATA2.
They are much lower than the value obtained with SUMER as well
   \citep[27.65\,\kms\ measured by][]{1999A&A...349..636T}.

   \par
What makes the NUBs interesting is that their intensities tend to be inversely correlated with their
   line-widths.
We note that a similar inverse correlation
   has also been found in sunspot plumes.
These are bright regions in the transition region above sunspots as viewed in spectral lines formed at 10$^{5}$\,--\,10$^{6}$\,K,
   often related to large coronal loops rooted in sunspot umbrae \citep{1974ApJ...193L.143F,1976ApJ...210..575F,1982ApJ...253..323R}.
The densities of sunspot plumes tend to be significantly lower than in plages,
   whereas the temperatures tend to be one or two magnitudes lower\,\citep{1974ApJ...193L.143F}.
They are dominated by downflows\,\citep{1982SoPh...77...77D,2001SoPh..198...89B,2003A&A...407L..29D}
   exceeding 25\,\kms\ at temperatures close to 250 000 K\,\citep{2001SoPh..198...89B}. A comprehensive review of sunspots and sunspot plumes can be found in \citet{2003A&ARv..11..153S}.
Using SUMER O\,{\sc v} observations, \citet{2003A&A...407L..29D} also reported that
   the non-thermal velocities in a bright sunspot plume are 5 to 10\,\kms\ lower than that in its neighboring regions.
In contrast to sunspot plumes, the NUBs reported in this work are more dynamic and smaller.
The possible relationship between NUBs and sunspot plumes is worth a dedicated investigation which is, however, beyond the scope
    of the present manuscript.
Nonetheless, it is certain that the NUBs we reported are located above sunspots and have similar line widths to sunspot materials.
Further investigations are therefore necessary for understanding the generation mechanisms that generate these NUBs and the roles
    they play in the local dynamics of ARs and mass cycling in sunspots.

\par
\acknowledgments
{\it Acknowledgments:}
 We thank the anonymous referee for the very constructive and practical comments and suggestions,
 Drs. Hui Tian and J. Chae for the useful discussions.
This research is supported by the 973 program 2012CB825601 and
 National Natural Science Foundation of China (41404135, 41274178, 41474150, 41174154, 41274176, and 41474149).
ZH thanks the Shandong provincial Natural Science Foundation (ZR2014DQ006)
   and the China Postdoctoral Science Foundation.
MM acknowledges the Leverhulme Trust.
IRIS is a NASA small explorer mission developed and operated by LMSAL with mission operations
    executed at NASA Ames Research center and major contributions to downlink communications
    funded by the Norwegian Space Center through an ESA PRODEX contract.
ZH and MM thank ISSI for supporting the team on ``Solar UV bursts -- a new insight to magnetic reconnection''.


\begin{thebibliography}{}
\expandafter\ifx\csname natexlab\endcsname\relax\def\natexlab#1{#1}\fi

\bibitem[{{Alpert} {et~al.}(2016){Alpert}, {Tiwari}, {Moore}, {Winebarger}, \&
  {Savage}}]{2016ApJ...822...35A}
{Alpert}, S.~E., {Tiwari}, S.~K., {Moore}, R.~L., {Winebarger}, A.~R., \&
  {Savage}, S.~L. 2016, \apj, 822, 35

\bibitem[{{Brynildsen} {et~al.}(2001){Brynildsen}, {Maltby}, {Fredvik},
  {Kjeldseth-Moe}, \& {Wilhelm}}]{2001SoPh..198...89B}
{Brynildsen}, N., {Maltby}, P., {Fredvik}, T., {Kjeldseth-Moe}, O., \&
  {Wilhelm}, K. 2001, \solphys, 198, 89

\bibitem[{{Chae} {et~al.}(1998){Chae}, {Sch{\"u}hle}, \&
  {Lemaire}}]{1998ApJ...505..957C}
{Chae}, J., {Sch{\"u}hle}, U., \& {Lemaire}, P. 1998, \apj, 505, 957

\bibitem[{{Cirtain} {et~al.}(2013){Cirtain}, {Golub}, {Winebarger}, {de
  Pontieu}, {Kobayashi}, {Moore}, {Walsh}, {Korreck}, {Weber}, {McCauley},
  {Title}, {Kuzin}, \& {Deforest}}]{2013Natur.493..501C}
{Cirtain}, J.~W., {Golub}, L., {Winebarger}, A.~R., {et~al.} 2013, \nat, 493,
  501

\bibitem[{{De Pontieu} {et~al.}(2015){De Pontieu}, {McIntosh},
  {Martinez-Sykora}, {Peter}, \& {Pereira}}]{2015ApJ...799L..12D}
{De Pontieu}, B., {McIntosh}, S., {Martinez-Sykora}, J., {Peter}, H., \&
  {Pereira}, T.~M.~D. 2015, \apjl, 799, L12

\bibitem[{{De Pontieu} {et~al.}(2009){De Pontieu}, {McIntosh}, {Hansteen}, \&
  {Schrijver}}]{2009ApJ...701L...1D}
{De Pontieu}, B., {McIntosh}, S.~W., {Hansteen}, V.~H., \& {Schrijver}, C.~J.
  2009, \apjl, 701, L1

\bibitem[{{De Pontieu} {et~al.}(2014){De Pontieu}, {Title}, {Lemen}, {Kushner},
  {Akin}, {Allard}, {Berger}, {Boerner}, {Cheung}, {Chou}, {Drake}, {Duncan},
  {Freeland}, {Heyman}, {Hoffman}, {Hurlburt}, {Lindgren}, {Mathur}, {Rehse},
  {Sabolish}, {Seguin}, {Schrijver}, {Tarbell}, {W{\"u}lser}, {Wolfson},
  {Yanari}, {Mudge}, {Nguyen-Phuc}, {Timmons}, {van Bezooijen}, {Weingrod},
  {Brookner}, {Butcher}, {Dougherty}, {Eder}, {Knagenhjelm}, {Larsen},
  {Mansir}, {Phan}, {Boyle}, {Cheimets}, {DeLuca}, {Golub}, {Gates}, {Hertz},
  {McKillop}, {Park}, {Perry}, {Podgorski}, {Reeves}, {Saar}, {Testa}, {Tian},
  {Weber}, {Dunn}, {Eccles}, {Jaeggli}, {Kankelborg}, {Mashburn}, {Pust},
  {Springer}, {Carvalho}, {Kleint}, {Marmie}, {Mazmanian}, {Pereira}, {Sawyer},
  {Strong}, {Worden}, {Carlsson}, {Hansteen}, {Leenaarts}, {Wiesmann},
  {Aloise}, {Chu}, {Bush}, {Scherrer}, {Brekke}, {Martinez-Sykora}, {Lites},
  {McIntosh}, {Uitenbroek}, {Okamoto}, {Gummin}, {Auker}, {Jerram}, {Pool}, \&
  {Waltham}}]{2014SoPh..289.2733D}
{De Pontieu}, B., {Title}, A.~M., {Lemen}, J.~R., {et~al.} 2014, \solphys, 289,
  2733

\bibitem[{{Deng} {et~al.}(2016){Deng}, {Yurchyshyn}, {Tian}, {Kleint}, {Liu},
  {Xu}, \& {Wang}}]{2016arXiv160700306D}
{Deng}, N., {Yurchyshyn}, V., {Tian}, H., {et~al.} 2016, ArXiv e-prints,
  arXiv:1607.00306

\bibitem[{{Dere}(1982)}]{1982SoPh...77...77D}
{Dere}, K.~P. 1982, \solphys, 77, 77

\bibitem[{{Dere} {et~al.}(1989){Dere}, {Bartoe}, \&
  {Brueckner}}]{1989SoPh..123...41D}
{Dere}, K.~P., {Bartoe}, J.-D.~F., \& {Brueckner}, G.~E. 1989, \solphys, 123,
  41

\bibitem[{{Dere} {et~al.}(1997){Dere}, {Landi}, {Mason}, {Monsignori Fossi}, \&
  {Young}}]{1997A&AS..125..149D}
{Dere}, K.~P., {Landi}, E., {Mason}, H.~E., {Monsignori Fossi}, B.~C., \&
  {Young}, P.~R. 1997, \aaps, 125, 149

\bibitem[{{Dere} \& {Mason}(1993)}]{1993SoPh..144..217D}
{Dere}, K.~P., \& {Mason}, H.~E. 1993, \solphys, 144, 217

\bibitem[{{Doyle} \& {Madjarska}(2003)}]{2003A&A...407L..29D}
{Doyle}, J.~G., \& {Madjarska}, M.~S. 2003, \aap, 407, L29

\bibitem[{{Foukal}(1976)}]{1976ApJ...210..575F}
{Foukal}, P.~V. 1976, \apj, 210, 575

\bibitem[{{Foukal} {et~al.}(1974){Foukal}, {Noyes}, {Reeves}, {Schmahl},
  {Timothy}, {Vernazza}, {Wilhbroe}, \& {Huber}}]{1974ApJ...193L.143F}
{Foukal}, P.~V., {Noyes}, R.~W., {Reeves}, E.~M., {et~al.} 1974, \apjl, 193,
  L143

\bibitem[{{Harrison} {et~al.}(1999){Harrison}, {Lang}, {Brooks}, \&
  {Innes}}]{1999A&A...351.1115H}
{Harrison}, R.~A., {Lang}, J., {Brooks}, D.~H., \& {Innes}, D.~E. 1999, \aap,
  351, 1115

\bibitem[{{Huang} {et~al.}(2014){Huang}, {Madjarska}, {Xia}, {Doyle},
  {Galsgaard}, \& {Fu}}]{2014ApJ...797...88H}
{Huang}, Z., {Madjarska}, M.~S., {Xia}, L., {et~al.} 2014, \apj, 797, 88

\bibitem[{{Huang} {et~al.}(2015){Huang}, {Xia}, {Li}, \&
  {Madjarska}}]{2015ApJ...810...46H}
{Huang}, Z., {Xia}, L., {Li}, B., \& {Madjarska}, M.~S. 2015, \apj, 810, 46

\bibitem[{{Innes} {et~al.}(1997){Innes}, {Inhester}, {Axford}, \&
  {Wilhelm}}]{1997Natur.386..811I}
{Innes}, D.~E., {Inhester}, B., {Axford}, W.~I., \& {Wilhelm}, K. 1997, \nat,
  386, 811

\bibitem[{{Kleint} {et~al.}(2014){Kleint}, {Antolin}, {Tian}, {Judge}, {Testa},
  {De Pontieu}, {Mart{\'{\i}}nez-Sykora}, {Reeves}, {Wuelser}, {McKillop},
  {Saar}, {Carlsson}, {Boerner}, {Hurlburt}, {Lemen}, {Tarbell}, {Title},
  {Golub}, {Hansteen}, {Jaeggli}, \& {Kankelborg}}]{2014ApJ...789L..42K}
{Kleint}, L., {Antolin}, P., {Tian}, H., {et~al.} 2014, \apjl, 789, L42

\bibitem[{{Kobayashi} {et~al.}(2014){Kobayashi}, {Cirtain}, {Winebarger},
  {Korreck}, {Golub}, {Walsh}, {De Pontieu}, {DeForest}, {Title}, {Kuzin},
  {Savage}, {Beabout}, {Beabout}, {Podgorski}, {Caldwell}, {McCracken},
  {Ordway}, {Bergner}, {Gates}, {McKillop}, {Cheimets}, {Platt}, {Mitchell}, \&
  {Windt}}]{2014SoPh..289.4393K}
{Kobayashi}, K., {Cirtain}, J., {Winebarger}, A.~R., {et~al.} 2014, \solphys,
  289, 4393

\bibitem[{{Kwak} {et~al.}(2016){Kwak}, {Chae}, {Song}, {Kim}, {Lim}, \&
  {Madjarska}}]{2016ApJ...821L..30K}
{Kwak}, H., {Chae}, J., {Song}, D., {et~al.} 2016, \apjl, 821, L30

\bibitem[{{Landi} {et~al.}(2013){Landi}, {Young}, {Dere}, {Del Zanna}, \&
  {Mason}}]{2013ApJ...763...86L}
{Landi}, E., {Young}, P.~R., {Dere}, K.~P., {Del Zanna}, G., \& {Mason}, H.~E.
  2013, \apj, 763, 86

\bibitem[{{O'Dwyer} {et~al.}(2010){O'Dwyer}, {Del Zanna}, {Mason}, {Weber}, \&
  {Tripathi}}]{2010A&A...521A..21O}
{O'Dwyer}, B., {Del Zanna}, G., {Mason}, H.~E., {Weber}, M.~A., \& {Tripathi},
  D. 2010, \aap, 521, A21

\bibitem[{{Parnell} {et~al.}(2002){Parnell}, {Bewsher}, \&
  {Harrison}}]{2002SoPh..206..249P}
{Parnell}, C.~E., {Bewsher}, D., \& {Harrison}, R.~A. 2002, \solphys, 206, 249

\bibitem[{{Pereira} {et~al.}(2015){Pereira}, {Carlsson}, {De Pontieu}, \&
  {Hansteen}}]{2015ApJ...806...14P}
{Pereira}, T.~M.~D., {Carlsson}, M., {De Pontieu}, B., \& {Hansteen}, V. 2015,
  \apj, 806, 14

\bibitem[{{Pereira} {et~al.}(2014){Pereira}, {De Pontieu}, {Carlsson},
  {Hansteen}, {Tarbell}, {Lemen}, {Title}, {Boerner}, {Hurlburt}, {W{\"u}lser},
  {Mart{\'{\i}}nez-Sykora}, {Kleint}, {Golub}, {McKillop}, {Reeves}, {Saar},
  {Testa}, {Tian}, {Jaeggli}, \& {Kankelborg}}]{2014ApJ...792L..15P}
{Pereira}, T.~M.~D., {De Pontieu}, B., {Carlsson}, M., {et~al.} 2014, \apjl,
  792, L15

\bibitem[{{Peter} {et~al.}(2013){Peter}, {Bingert}, {Klimchuk}, {de Forest},
  {Cirtain}, {Golub}, {Winebarger}, {Kobayashi}, \&
  {Korreck}}]{2013A&A...556A.104P}
{Peter}, H., {Bingert}, S., {Klimchuk}, J.~A., {et~al.} 2013, \aap, 556, A104

\bibitem[{{Peter} {et~al.}(2014){Peter}, {Tian}, {Curdt}, {Schmit}, {Innes},
  {De Pontieu}, {Lemen}, {Title}, {Boerner}, {Hurlburt}, {Tarbell}, {Wuelser},
  {Mart{\'{\i}}nez-Sykora}, {Kleint}, {Golub}, {McKillop}, {Reeves}, {Saar},
  {Testa}, {Kankelborg}, {Jaeggli}, {Carlsson}, \&
  {Hansteen}}]{2014Sci...346C.315P}
{Peter}, H., {Tian}, H., {Curdt}, W., {et~al.} 2014, Science, 346, C315

\bibitem[{{Raymond} \& {Foukal}(1982)}]{1982ApJ...253..323R}
{Raymond}, J.~C., \& {Foukal}, P. 1982, \apj, 253, 323

\bibitem[{{R{\'e}gnier} {et~al.}(2014){R{\'e}gnier}, {Alexander}, {Walsh},
  {Winebarger}, {Cirtain}, {Golub}, {Korreck}, {Mitchell}, {Platt}, {Weber},
  {De Pontieu}, {Title}, {Kobayashi}, {Kuzin}, \&
  {DeForest}}]{2014ApJ...784..134R}
{R{\'e}gnier}, S., {Alexander}, C.~E., {Walsh}, R.~W., {et~al.} 2014, \apj,
  784, 134

\bibitem[{{Rouppe van der Voort} {et~al.}(2015){Rouppe van der Voort}, {De
  Pontieu}, {Pereira}, {Carlsson}, \& {Hansteen}}]{2015ApJ...799L...3R}
{Rouppe van der Voort}, L., {De Pontieu}, B., {Pereira}, T.~M.~D., {Carlsson},
  M., \& {Hansteen}, V. 2015, \apjl, 799, L3

\bibitem[{{Solanki}(2003)}]{2003A&ARv..11..153S}
{Solanki}, S.~K. 2003, \aapr, 11, 153

\bibitem[{{Teriaca} {et~al.}(1999){Teriaca}, {Banerjee}, \&
  {Doyle}}]{1999A&A...349..636T}
{Teriaca}, L., {Banerjee}, D., \& {Doyle}, J.~G. 1999, \aap, 349, 636

\bibitem[{{Testa} {et~al.}(2013){Testa}, {De Pontieu},
  {Mart{\'{\i}}nez-Sykora}, {DeLuca}, {Hansteen}, {Cirtain}, {Winebarger},
  {Golub}, {Kobayashi}, {Korreck}, {Kuzin}, {Walsh}, {DeForest}, {Title}, \&
  {Weber}}]{2013ApJ...770L...1T}
{Testa}, P., {De Pontieu}, B., {Mart{\'{\i}}nez-Sykora}, J., {et~al.} 2013,
  \apjl, 770, L1

\bibitem[{{Tian} {et~al.}(2014){Tian}, {Kleint}, {Peter}, {Weber}, {Testa},
  {DeLuca}, {Golub}, \& {Schanche}}]{2014ApJ...790L..29T}
{Tian}, H., {Kleint}, L., {Peter}, H., {et~al.} 2014, \apjl, 790, L29

\bibitem[{{Tian} {et~al.}(2011){Tian}, {McIntosh}, {De Pontieu},
  {Mart{\'{\i}}nez-Sykora}, {Sechler}, \& {Wang}}]{2011ApJ...738...18T}
{Tian}, H., {McIntosh}, S.~W., {De Pontieu}, B., {et~al.} 2011, \apj, 738, 18

\bibitem[{{Tsiropoula} {et~al.}(2012){Tsiropoula}, {Tziotziou}, {Kontogiannis},
  {Madjarska}, {Doyle}, \& {Suematsu}}]{2012SSRv..169..181T}
{Tsiropoula}, G., {Tziotziou}, K., {Kontogiannis}, I., {et~al.} 2012, \ssr,
  169, 181

\bibitem[{{Vissers} {et~al.}(2015){Vissers}, {Rouppe van der Voort}, \&
  {Carlsson}}]{2015ApJ...811L..33V}
{Vissers}, G.~J.~M., {Rouppe van der Voort}, L.~H.~M., \& {Carlsson}, M. 2015,
  \apjl, 811, L33

\bibitem[{{Wilhelm}(2000)}]{2000A&A...360..351W}
{Wilhelm}, K. 2000, \aap, 360, 351

\bibitem[{{Winebarger} {et~al.}(2013){Winebarger}, {Walsh}, {Moore}, {De
  Pontieu}, {Hansteen}, {Cirtain}, {Golub}, {Kobayashi}, {Korreck}, {DeForest},
  {Weber}, {Title}, \& {Kuzin}}]{2013ApJ...771...21W}
{Winebarger}, A.~R., {Walsh}, R.~W., {Moore}, R., {et~al.} 2013, \apj, 771, 21

\bibitem[{{Xia}(2003)}]{xiathesis}
{Xia}, L. 2003, {Equatorial Coronal Holes and Their Relation to the High-Speed
  Solar Wind Streams} (Ph.D. Thesis, G\"ottingen: Georg-August-Univ.)

\end{thebibliography}
\end{document}